\documentclass[aps,prl,superscriptaddress,reprint]{revtex4-1}
\usepackage{amsmath}
\usepackage{amssymb}
\usepackage{graphicx}
\usepackage[colorlinks,urlcolor=blue]{hyperref}
\usepackage{url}
\usepackage{color,xcolor}
\usepackage{ulem}
\usepackage{float}
\usepackage{epstopdf}
\begin{document}

\title{Peierls transition, ferroelectricity, and spin-singlet formation in the monolayer VOI$_2$}
\author{Yang Zhang}
\author{Ling-Fang Lin}
\affiliation{Department of Physics and Astronomy, University of Tennessee, Knoxville, TN 37996, USA}
\author{Adriana Moreo}
\affiliation{Department of Physics and Astronomy, University of Tennessee, Knoxville, TN 37996, USA}
\affiliation{Materials Science and Technology Division, Oak Ridge National Laboratory, Oak Ridge, TN 37831, USA}
\author{Gonzalo Alvarez}
\affiliation{Computational Sciences \& Engineering Division and Center for Nanophase Materials Sciences, Oak Ridge National Laboratory, Oak Ridge, TN 37831, USA}
\author{Elbio Dagotto}
\affiliation{Department of Physics and Astronomy, University of Tennessee, Knoxville, TN 37996, USA}
\affiliation{Materials Science and Technology Division, Oak Ridge National Laboratory, Oak Ridge, TN 37831, USA}

\date{\today}

\begin{abstract}
Using {\it ab initio} density functional theory and single-orbital Hubbard model calculations via the density matrix renormalization group method, we systematically studied the monolayer VOI$_2$ with a $3d^1$ electronic configuration. Our phonon calculations indicate that the orthorhombic $Pmm2$ FE-II phase is the most likely ground state, involving a ferroelectric distortion along the $a$-axis and V-V dimerization along the $b$-axis. Specifically, the ``pseudo Jahn-Teller'' effect caused by the coupling between empty V ($d_{xz/yz}$ and $d_{3z^2-r^2}$) and O $2p$ states is proposed as the mechanism that stabilizes the ferroelectric distortion from the paraelectric phase. Moreover, the half-filled metallic $d_{xy}$ band displays a Peierls instability along the $b$-axis, inducing a V-V dimerization. We also found very short-range antiferromagnetic coupling along the V-V chain due to the formation of nearly-decoupled spin singlets in the ground state.

\end{abstract}

\maketitle

\textit{Introduction.-}
Low-dimensional systems have attracted considerable interest for decades because their interactions between transition metals are strongly enhanced through electron-electron, phonon-phonon, electron-phonon and spin-phonon couplings, leading to rich physical properties~\cite{Monceau:ap,Dagotto:rmp94,Dagotto:Rmp,Grioni:JPCM,Lin:prl}. As the simplest systems, one-dimensional (1D) chains and ladders display remarkable states, potentially important for
applications~\cite{Grioni:JPCM,cu-ladder1,Zhang:prb19,Zhang:prb20-2,gao:prb20}. For example, considering electronic correlation effects, the Cu- and Fe-based ladders become superconducting under pressure {\color{blue}(at $12$ K in Cu-based ladder and $24$ K in Fe-based ladders).} ~\cite{cu-ladder2,cu-ladder3,Takahashi:Nm,Zhang:prb17,Ying:prb17,Zhang:prb18} Chains often undergo Peierls phase transitions induced by strong electron-phonon coupling {\color{blue}(e.g. (TaSe$_4$)$_2$I at $263$ K)}~\cite{Gooth:nature,Zhang:prb20-1}. Due to the empty W-$d^0$ orbital, WOX$_4$ halogens were predicted to be ferroelectric~\cite{Lin:prm} {\color{blue}above room temperature}. Considering the spin-phonon interaction, multiferroelectric behavior was also expected in some 1D systems~\cite{Choi:prl,Zhang:prb20-2}.

Recently, the monolayer VOI$_2$, with a $d^1$ configuration ($S=1/2$), was predicted to be {\color{blue}multiferroic}~\cite{Tan:prb,Ding:prb}. In addition, the electric-field switch of the magnetic topological charge was also realized in this system~\cite{Xu:prl}. However, there are still issues remaining to be addressed in the VOI$_2$ monolayer.  For example, in a $d^0$ system, such as BaTiO$_3$, the large ferroelectric polarization can be explained by the so-called ``pseudo Jahn-Teller'' (p-JT) effect, which reduces the total energy through a noncentrosymmetric distortion~\cite{Cohen:nature,Cohen:fe,Young:prb}. However, for the $d^1$ VOI$_2$ material, why a large ferroelectric distortion is stable? Moreover, the half-filled metallic $d_{xy}$ band in the undistorted VI$_2$ chain along the $b$-axis should be unstable according to Peierls' theorem ~\cite{Smaalen:aca}. In other words, both dimerization and a metal-insulator transition (MIT) are expected to occur in the metallic $S=1/2$ VI$_2$ chain at low temperatures. To our knowledge, these intriguing questions remain unexplored.

To better understand these issues, here both the density functional theory (DFT) and density matrix renormalization group (DMRG) methods are employed to investigate the monolayer VOI$_2$ in more detail. {\color{blue}The DFT calculations were performed based on the projector augmented wave (PAW) method with the Perdew-Burke-Ernzerhof (PBE) exchange potential, as implemented in the Vienna {\it ab initio} simulation package (VASP) code~\cite{Kresse:Prb,Kresse:Prb96,Blochl:Prb}.} First, we found that
the Peierls transition indeed occurs along the $b$-axis in this system, resulting in a V-dimerized chain
and concomitant MIT. Second, we observed that
the p-JT effect caused by the coupling between the empty V-$3d$ orbitals $d_{xz/yz}$ and $d_{3z^2-r^2}$, in
combination with the O-$2p$ orbitals, stabilizes the ferroelectric distortion along the $a$-axis. Third, based on
DMRG calculations ~\cite{dmrgcontext}, we found that this system has antiferromagnetic-antiferromagnetic coupling along the dimerized
VI$_2$ chain, although without long-range magnetic ordering.

\begin{figure}
\centering
\includegraphics[width=0.48\textwidth]{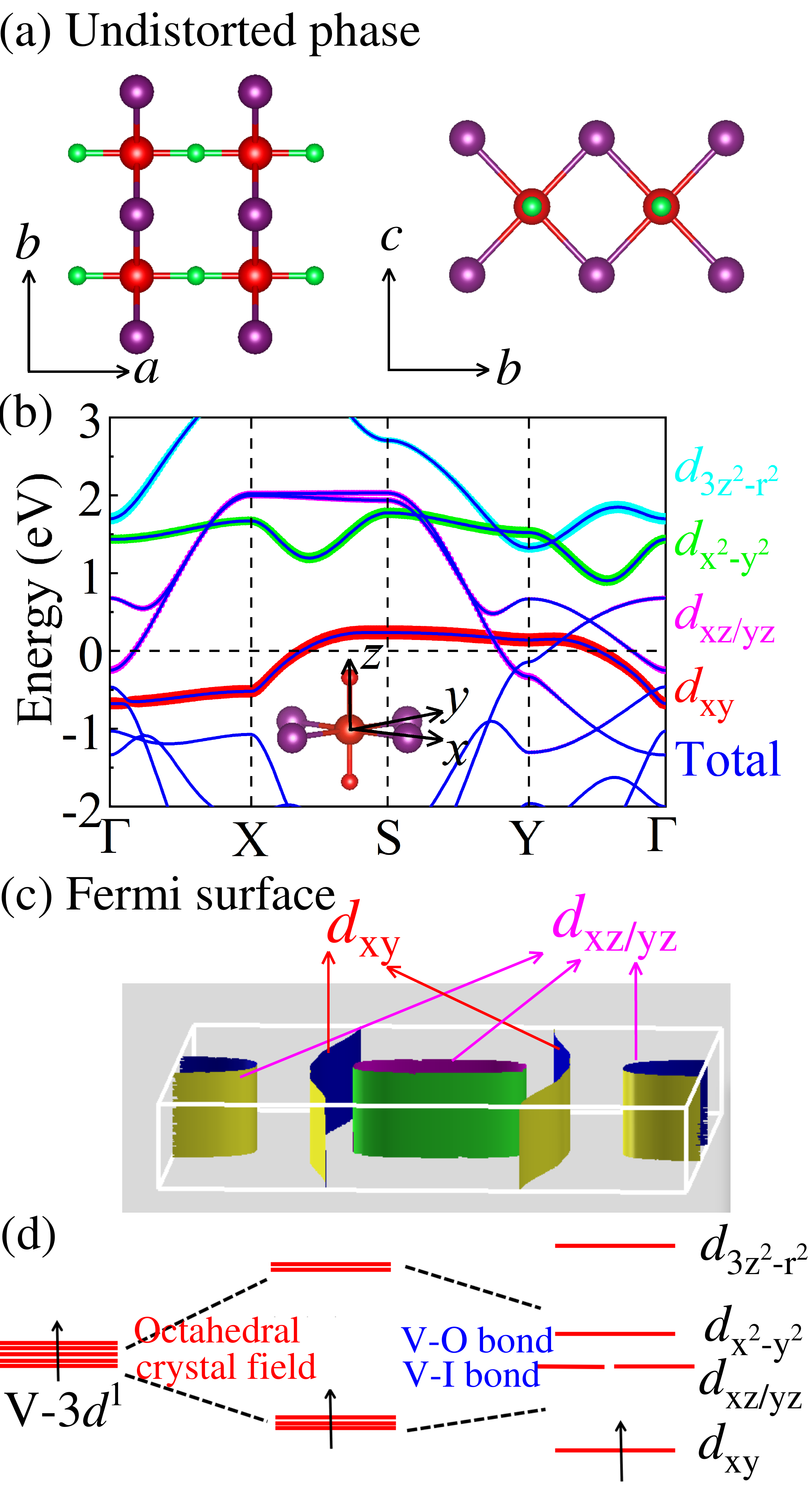}
\caption{(a) Schematic crystal structure of the VOI$_2$ conventional cell (red = V; green = O; purple = I). (b) Projected band structures of the undistorted monolayer VOI$_2$ for the NM state. The Fermi level is shown with dashed lines. The weight of each vanadium orbital is represented by the size of the circles. The coordinates of the high symmetry points in the plane Brillouin zone (BZ) are  $\Gamma$ = (0, 0, 0), X = (0.5, 0, 0), S = (0.5, 0.5, 0), Y = (0, 0.5, 0). (c)  Fermi surface. (d) The energy splitting of V's 3$d$ orbitals with the $d^1$ configuration.}
\label{Fig1}
\end{figure}

\textit{Undistorted phase.-}
The undistorted monolayer VOI$_2$ is in an orthorhombic crystal structure with space group Pmmm (No.47), where the VO$_2$I$_4$ octahedra form a two-dimensional plane. These VO$_2$I$_4$ octahedra are corner-sharing and edge-sharing along the $a$- and $b$-axis, respectively. As shown in Fig.~\ref{Fig1}(a), this system contains a VI$_2$ chain along the $b$-axis with identical V-I bonds (and VO chains along the $a$-axis). Before addressing the structural instability, let us discuss the electronic structure corresponding to the non-magnetic (NM) state of the undistorted VOI$_2$.

As shown in Fig.~\ref{Fig1}(b), the $e_g$ orbitals $d_{x^2-y^2}$ and $d_{3z^2-r^2}$ are located at high energy and, thus, unoccupied. The Fermi surface is formed by the mainly occupied $d_{xy}$ orbitals and partially occupied degenerate itinerant $d_{xz/yz}$ orbitals. {\color{blue}Note that the $d_{xy}$ orbital lays on the $bc$ plane, with the $x$ or $y$ axis along V-I directions and the $z$ axis being the $a$-axis [see Fig.~\ref{Fig1}(b)].} It is also clearly shown that the V $d_{xy}$ band is much more dispersive along the $b$-axis (X-S or Y-$\Gamma$ paths) than the $a$-axis ($\Gamma$-X or S-Y paths), strongly suggesting quasi-one-dimensional electronic itineracy along the $b$-axis. Accordingly, the Fermi surface shown in Fig.~\ref{Fig1}(c)  indicates that the $d_{xy}$ band has 1D behavior while the $d_{xz/yz}$ bands display 2D behavior.

Furthermore, the energy splitting of the vanadium $d$ orbitals with the $d^1$ configuration is sketched in Fig.~\ref{Fig1}(d). The octahedral crystal field leads to three lower energy $t_{2g}$ orbitals ($d_{xy}$,
$d_{yz}$, and $d_{xz}$) and two higher energy $e_g$ orbitals ($d_{x^2-y^2}$ and $d_{3z^2-r^2}$). In addition, the replacement of the I atom by the O atom at the octahedral apex induce two different V-$X$ ($X=$ O or I) bonds with two shorten V-O bonds along the $a$-axis ($z$-axis) and four enlongated V-I bonds along the $b-c$ ($xy$) plane, resulting
in the $d_{xy}$ energy level shifting down, lower than that the  $d_{yz}$ and $d_{xz}$ levels.

\textit{Structural instability.-}
In a $S=1/2$ one-dimensional chain, the system is not stable at low temperature because a structural distortion usually occurs along the chain direction through strong electron-phonon interactions, the so-called Peierls transition.

\begin{figure}
\centering
\includegraphics[width=0.48\textwidth]{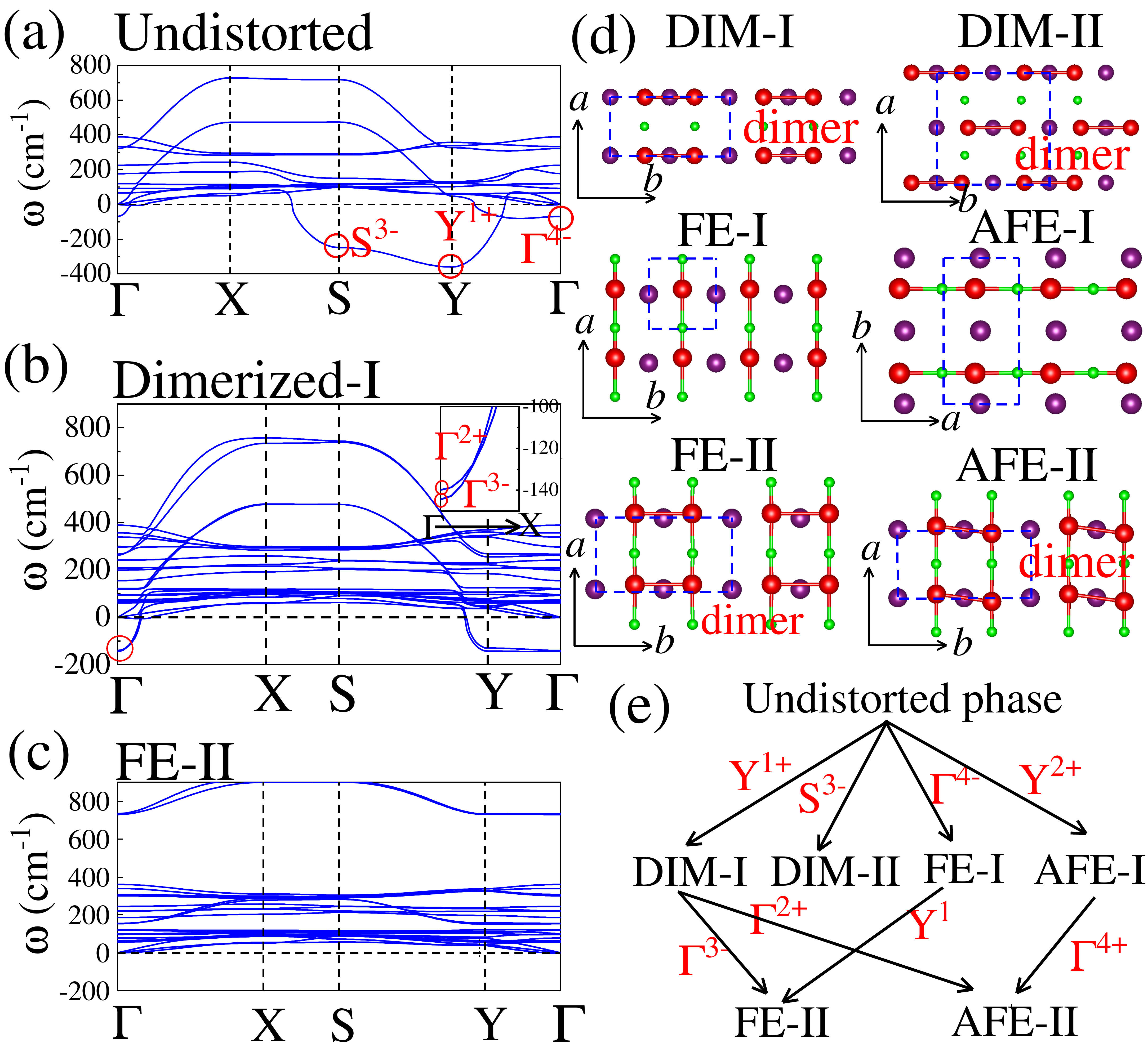}
\caption{(a-c) The phonon spectrum of VOI$_2$ for the undistorted, DIM-I, and FE-II phases. Here, the $4\times4\times1$ and $4\times2\times1$ supercell was selected for the undistorted and DIM-I/FE-II phases in our calculations with the non-magnetism state, respectively. (a) Undistorted, (b) Dimmerized-I, and (c) FE-II cases. (d) Sketch of the possible structural patterns studied here. (e) The group theory analysis for the monolayer VOI$_2$.}
\label{Fig2}
\end{figure}

To better understand the structural phase transition in VOI$_2$, we performed the phononic dispersion calculations using a $4\times4\times1$ supercell for the undistorted phase, as shown in Fig.~\ref{Fig2}(a). The phonon spectra indicates the presence of three imaginary frequencies appearing at the $\Gamma$, S, and Y points of the undistorted structure, respectively. According to group theory analysis using the AMPLIMODES software~\cite{Orobengoa:jac,Perez-Mato:aca}, these spontaneous distortion modes are the $\Gamma^{4-}$, S$^{3-}$, and Y$^{1+}$ modes, respectively~\cite{Modecontext}.

Next, we fully relaxed the crystal lattice from the original undistorted phase along those three mode displacements, and then obtained three phases, namely DIM-I, DIM-II, and FE-I phases. Based on the relaxed structures, we found that the DIM-I phase with a V-V dimerization along the $b$-axis has the lowest energy among those three configurations (see Table~\ref{Table1}). As expected, a MIT is induced by the V-V dimerization, resulting in a Peierls transition
that opens a gap.

Next, we calculated the phononic dispersion spectrum for the DIM-I phase, finding two imaginary frequency modes, as shown in Fig.~\ref{Fig2}(b), corresponding to the $\Gamma^{2+}$ and $\Gamma^{3-}$ distortions, respectively. By extracting those two unstable phononic modes of the DIM-I phase, antiferroelectric dimerized (AFE-II) and ferroelectric dimerized (FE-II) phases were obtained, see Fig.~\ref{Fig2}(d). After a full lattice relaxation for both the FE-II and AFE-II configurations, the FE-II state was found to have a lower energy than the AFE-II state (by $\sim$ 7.8 meV/V).  Comparing with other structural configurations (see Table~\ref{Table1}), the FE-II phase has the lowest energy overall. In addition, Fig.~\ref{Fig2}(c) indicates that the FE-II phase is now dynamically stable since no  additional imaginary frequency modes were obtained in the phononic dispersion spectrum.

Furthermore, based on the AMPLIMODES software \cite{Orobengoa:jac,Perez-Mato:aca}, we also performed the group theory analysis for the monolayer VOI$_2$, as shown in Fig.~\ref{Fig2}(e). The FE-II phase can be regarded as a combination between the DIM-I (V-V dimerization along $b$-axis) and FE-I (ferroelectric distortion along $a$-axis) modes. Additional DFT results for the FE-I and DIM-I phases are reported in the Supplementary Material (SM)~\cite{Supplemental}.

\begin{table}
\centering\caption{The optimized
lattice constants ({\AA}), and band gaps (eV) for many configurations, as well as the energy differences (meV/V) with respect to the undistorted configuration, taken as the reference of energy.}
\begin{tabular*}{0.48\textwidth}{@{\extracolsep{\fill}}lllc}
\hline
\hline
  & $a$/$b$ &  Gap  & Energy \\
\hline
Undistorted phase      & 3.6528/3.7818  & 0    &  0   \\
DIM-I      & 3.6504/7.4181 & 0.14 &   -201.6 \\
DIM-II     & 7.2674/7.4424 & 0.10 &  -79.1 \\
FE-I       & 3.8317/3.7489 & 0 & -30.9 \\
AFE-I      & 3.7963/7.5385 & 0 & -5.1  \\
FE-II      & 3.8137/7.3446  &0.53 & -238.5  \\
AFE-II      & 3.7866/7.3529 & 0.43 & -230.7 \\
\hline
\hline
\end{tabular*}
\label{Table1}
\end{table}

\textit{Pseudo Jahn-Teller effect.-}
In some non-$d^0$ perovskite cases~\cite{Filippetti:prb,Rondinelli:prb}, the system could still
undergo ``p-JT'' off-centering distortions, stabilizing the polar ground state. According to the previous analysis of the VO$_2$I$_4$ octahedra, V$^{\rm 4+}$ has only one electron placed in the $d_{xy}$ orbital, resulting in empty {\color{blue}$d_{xz/yz}$ and $d_{3z^2-r^2}$ orbitals.}

\begin{figure*}
\centering
\includegraphics[width=0.96\textwidth]{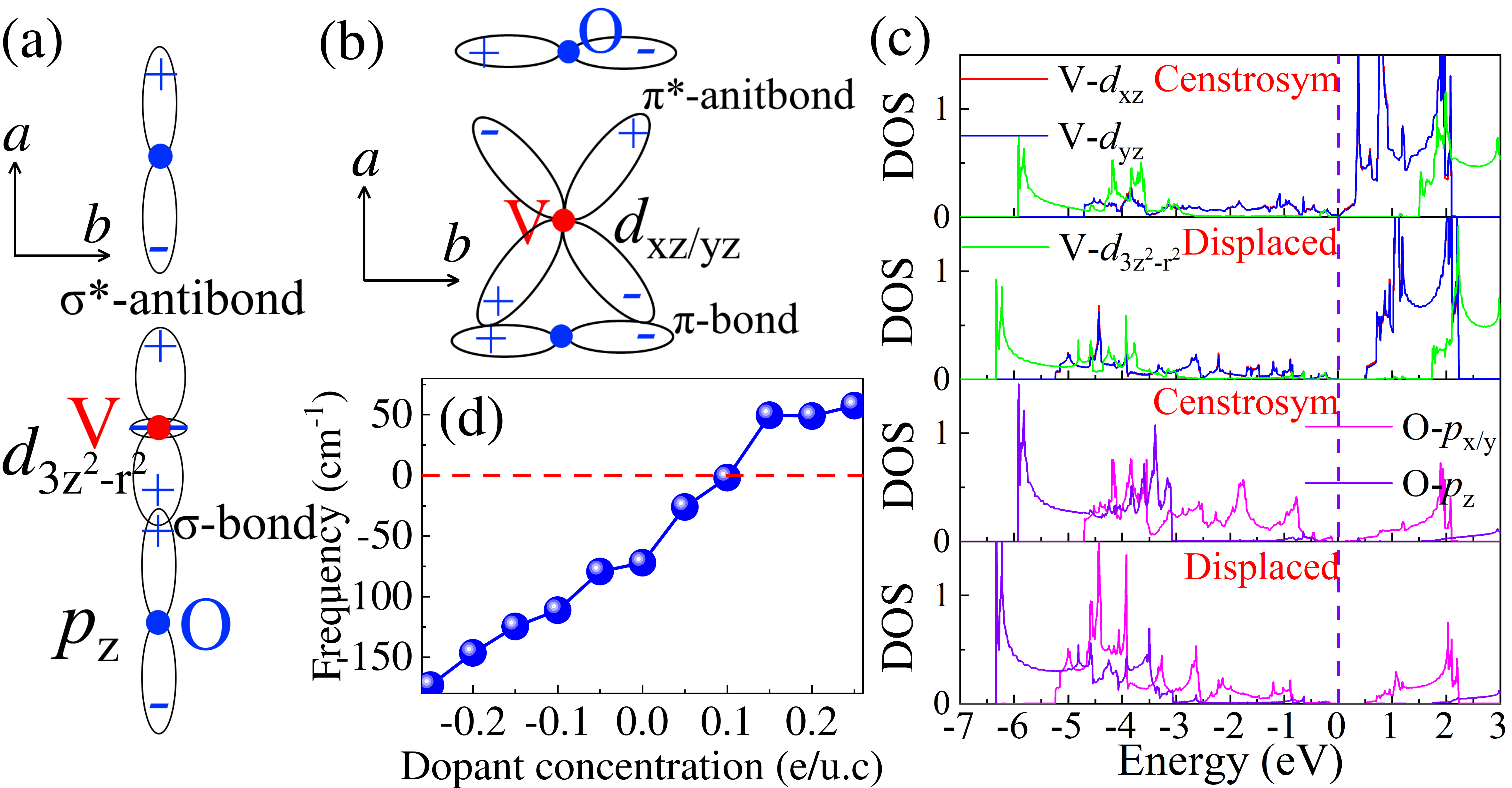}
\caption{(a-b) Schematics of the empty $p-d$ hybridizations in the ferroelectric monolayer VOI$_2$ at the $\Gamma$ point. Here, the colored solid dots denote V (red) and O (blue). (a) V($3d_{3z^2-r^2}$)-O($2p_z$) hybrid state. (b) V($3d_{xz/yz}$)-O($2p_x$) hybrid state. (c) Orbital-resolved density of V $3d$($d_{xz/yz}$ and $d_{3z^2-r^2}$) and O $2p$ ($p_x$ and $p_z$) states in the FE-II phase of monolayer VOI$_2$. ``Centrosym'' refers to the atoms in the centrosymmetric position while ``displaced'' indicates the atoms present ferroelectric displacements along the V-O bond ([100]). (d) Ferroelectric distortion mode ($\Gamma^{4-}$) obtained from the undistorted phase of monolayer VOI$_2$ vs. doping.}
\label{Fig3}
\end{figure*}

As shown in Fig.~\ref{Fig3}(a), the hybridization between empty V-$3d_{3z^2-r^2}$ and O-$2p_z$ orbitals leads to alternating $\sigma$-bonding and $\sigma^*$-antibonding states along the V-O chain. Furthermore, the V-$3d_{xz/yz}$ and O-$2p_{x/y}$ orbitals hybridization results from alternating $\pi$-bonding and $\pi^*$-antibonding states along the same V-O chain. As a result, the symmetry is broken along the V-O chain when the ferroelectric distortion occurs.

To better understand the stabilization of the ferroelectric distortion, we calculated the electronic structures for the centrosymmetric and  ferroelectric symmetry. As shown in Fig.~\ref{Fig3}(c), by comparing the density-of-states (DOS) with centrosymmetric and displaced positions, the energy of the valence band (VB) lowers by $0.4$ eV, while the conduction band (CB) levels move upward. These changes in the electronic structure support the ``p-JT'' effect caused by the interaction between empty $3d$ and occupied $2p$ states through the lattice distortion. The changes of energy levels of the VB and CB can reduce the total energy and stabilize the lattice distortion from a high-symmetric phase, resulting in a ferroelectric phase~\cite{pJTcontext}. Next, we also calculated the Born effective charge (BEC) of V along the [100] direction ($a$-axis) for the undistorted ($\sim$ 13.06), DIM-I ($\sim$ 13.43), FE-I ($\sim$ 4.27), and Fe-II ($\sim$ 4.84) states, respectively, which is consistent with previous calculations~\cite{Tan:prb}. {\color{blue}Moreover, we also calculated the BEC of O along the [100] direction ($a$-axis) for the undistorted ($\sim$ -11.86), DIM-I ($\sim$ -12.65), FE-I ($\sim$ -4.32), and Fe-II ($\sim$ -4.88) states, respectively.} The large anomalous deviations in the BECs from the formal charge (4 for V$^{\rm 4+}$) in the centrosymmetric phases, suggest a ferroelectric instability. In the displayed ferroelectric phases, the reduction of BECs also reflected the stabilization of ferroelectric distortion. In this case, the ferroelectric distortion can be stabilized by the ``p-JT'' effect caused by the coupling between empty $3d$ and occupied $p$ states.

To address the polar distortion instability, we also employed the virtual crystal approximation
(VCA)~\cite{VCAcontext} to investigate the effect of doping, which is widely used in the electronic structure context~\cite{Zhang:prb20,Bellaiche:Prb,Ramer:Prb,Zhao:prb18}. Under hole doping, we found that the ferroelectric instability is enhanced  because the doping initially affects the V-sites $d_{xy}$ orbitals that do not affect the hybridization between empty V-$3d$ and $O-2p$. However, under electron doping the ferroelectric (FE) distortion is reduced  due to the electronic occupation of $d_{xz/yz}$. This different doping behaviour is compatible with the notion that the FE distortion
is caused by the p-JT effect.

\textit{FE-II phase.-}
Based on the optimized crystal structure, the FE-II phase of monolayer VOI$_2$ is orthorhombic with space group $Pmm2$ (No.25) and ferroelectric polarization along the $a$-axis (V-O direction)~\cite{FE2context}. As discussed before, the FE-II phase can be regarded as the combination of the FE distortion ($\Gamma^{\rm 4-}$ mode) along the $a$-axis and V-V dimerization (Y$^{1+}$ mode) of the undistorted phase.

\begin{figure} [H]
\centering
\includegraphics[width=0.48\textwidth]{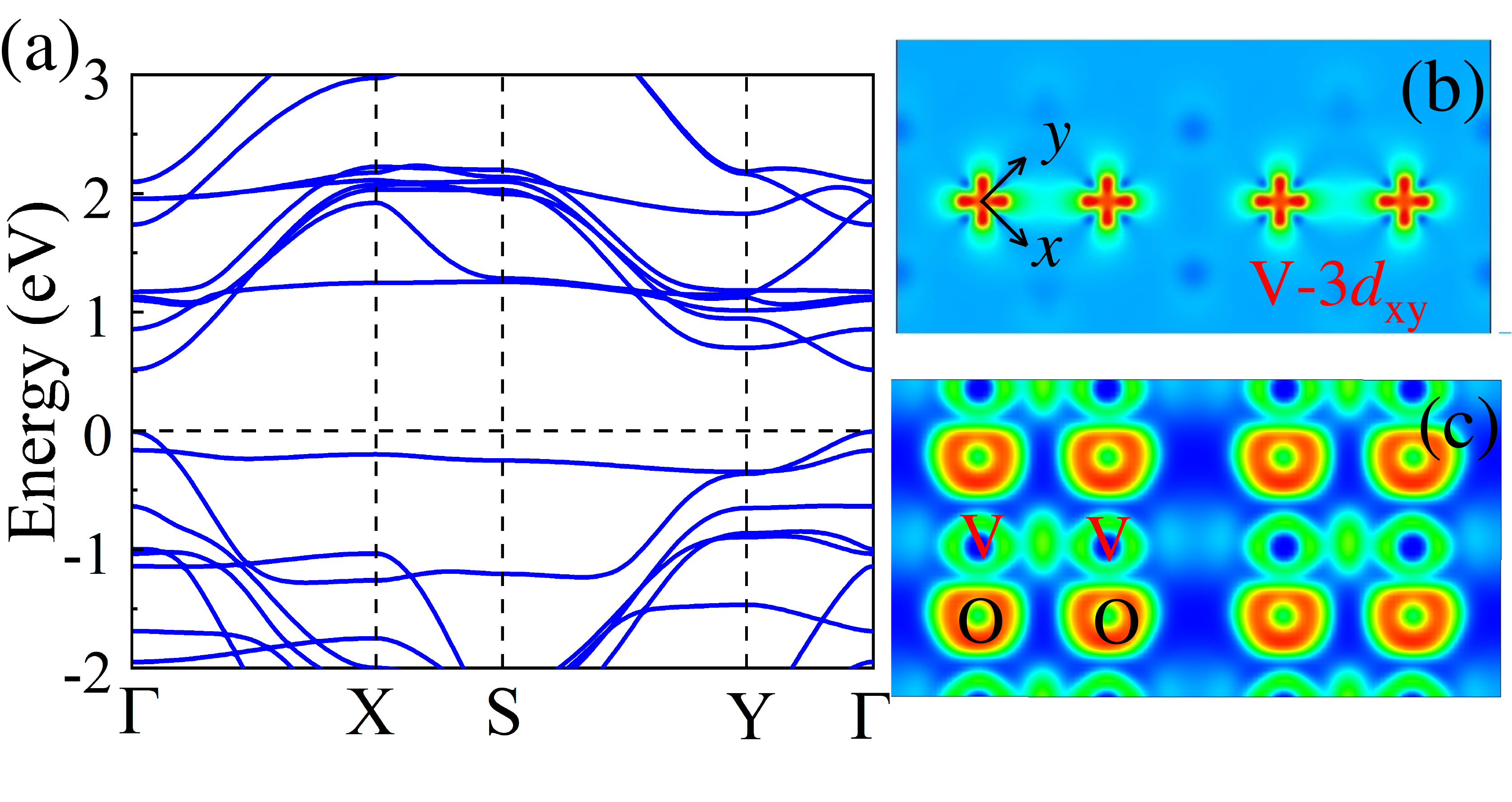}
\caption{(a) Band structure of the FE-II monolayer VOI$_2$ using the NM state. (b) The electronic density of the valence band near the Fermi level (-0.5 to 0 eV ) in the $bc$ plane, showing clear $d_{\rm xy}$ characteristics.
(c) Electron localization function of the FE-II monolayer VOI$_2$ in the $ab$ plane.}
\label{Fig4}
\end{figure}

Figure~\ref{Fig4}(a) shows that the $d_{xy}$ band is anisotropic, being more dispersive along the $b$-axis (Y-$\Gamma$ path) than the $a$-axis ($\Gamma$-X path). As shown in Fig.~\ref{Fig4}(b), along the $b$-axis the V-V dimerization distortion induces a dominant $d_{xy}$-$d_{xy}$ $\sigma$-bonding state, leading to a large overlap of $d_{xy}$ orbitals in the V-V dimer. Furthermore, we calculated the electron localization function (ELF)~\cite{Savin:Angewandte} for the FE-II phase that suggest covalent characteristics for the V-O bonds, supporting the previous analysis of the p-JT effect. Based on the Berry phase method~\cite{King-Smith:Prb,Resta:Rmp}, we estimate that the FE polarization ($P$) of the FE-II phase of monolayer VOI$_2$ is about $257$ pC/m. More details for FE-II are available in the SM~\cite{Supplemental}.

\begin{figure}
\centering
\includegraphics[width=0.48\textwidth]{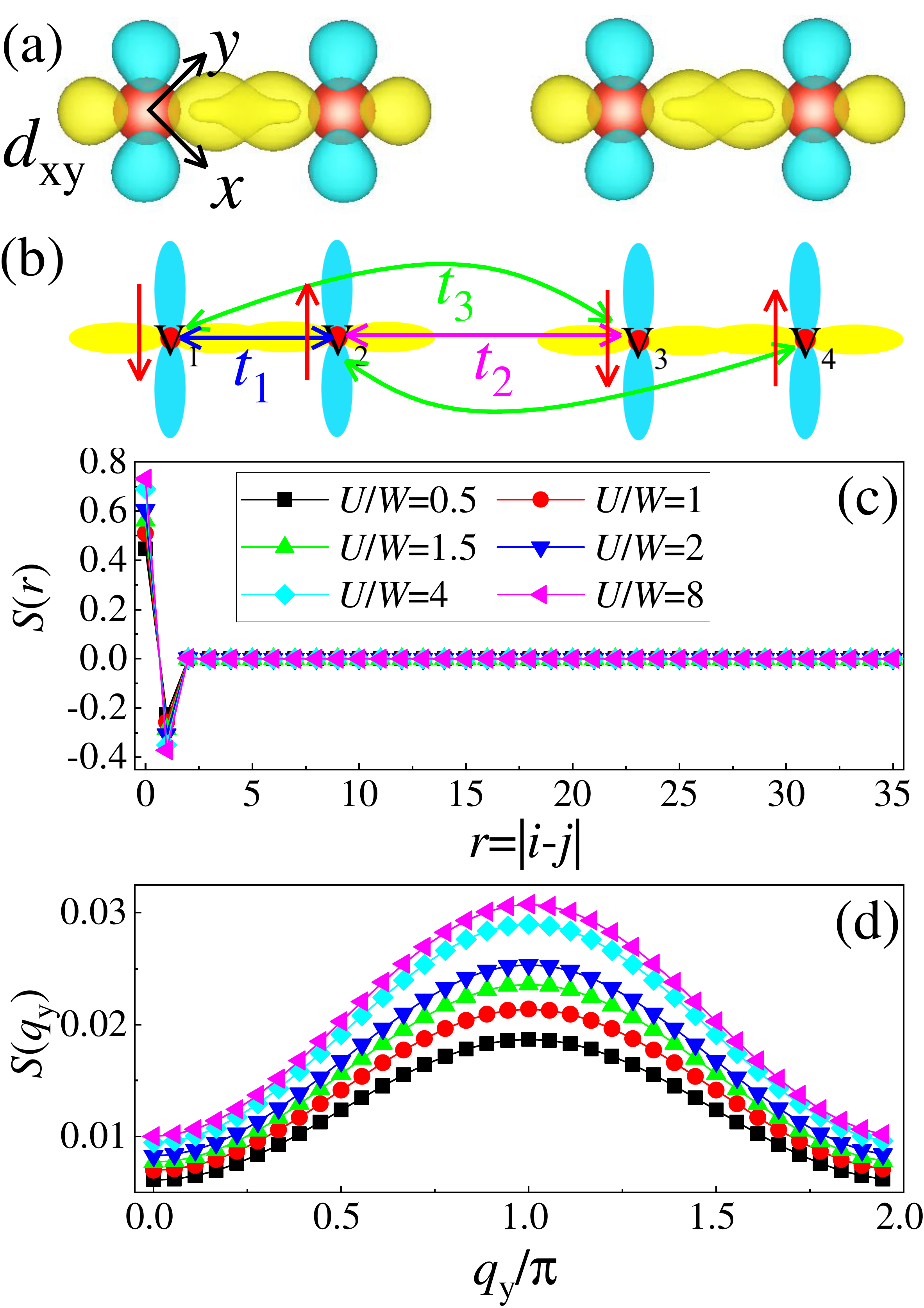}
\caption{(a) Wannier function of the V $3d_{xy}$ orbital, indicating the alternating $dd$$\sigma$-bonding and $dd$$\sigma^{*}$-antibonding states along the $b$-axis. (b) Different hoppings along the chain used in our DMRG calculations ($t_1=-0.647$ eV, $ t_2=0.067$ eV and $t_3=0.009$ eV). All the hopping amplitudes were obtained from the DFT calculations via Wannier functions. The spin arrows represent short-range antiferromagnetic-antiferromagnetic coupling along the chain. (c) S(r)=$\langle{{\bf S}_i \cdot {\bf S}_j}\rangle$ (with r=$\left|{i-j}\right|$) and (d) $S(q_y)$ for different values of $U/W$. Here, we used a $36$-sites cluster chain with two nearest-neighbors and one next-nearest-neighbor hoppings. (d) Spin structure factor at the six values of $U/W$ of panel (c).}
\label{Fig5}
\end{figure}

\textit{Magnetism-}
Based on the maximally localized Wannier method~\cite{Mostofi:cpc}, the Wannier function of the $3d_{xy}$ orbital was plotted in Fig.~\ref{Fig5}(a). It displays a strong $dd\sigma$-bonding state along the $b$-chain, suggesting the formation of a local singlet spin dimer. To better understand the magnetic coupling along this chain, an effective single-orbital Hubbard model was constructed to calculate the real-space spin correlations via the density matrix renormalization group (DMRG) method ~\cite{white:prl,white:prb}, where we have used the DMRG++ software~\cite{Alvarez:cpc}. The model studied here includes the kinetic energy and interaction energy terms $H = H_k + H_{int}$:

\begin{eqnarray}
H = \sum_{i,\sigma,{\alpha}}t_{{\alpha}}
(c^{\dagger}_{i\sigma}c^{\phantom\dagger}_{i+{\alpha},\sigma}+H.~c.)+ U\sum_in_{i\uparrow} n_{i\downarrow},
\end{eqnarray}
where the first term represents the hopping of an electron from site $i$ to site $i+{\alpha}$. The number ${\alpha}$ indicates the three different hopping paths shown in Fig.~\ref{Fig5}(b). The second term is the standard intraorbital Hubbard repulsion.

Figure~\ref{Fig5}(c) shows the spin-spin correlation $S(r)=\langle{{\bf S}_i \cdot {\bf S}_j}\rangle$ vs. distance $r$ for different values of $U/W$. The distance is $r=\left|{i-j}\right|$, with $i$ and $j$ site indexes. The spin-spin correlation decays very fast with distance $r$, suggesting a long-range disordered phase in this dimerized chain composed of strong dimer spin-singlet states ($(|{\uparrow \downarrow}\rangle -|{\downarrow \uparrow}\rangle)/\sqrt{2}$) nearly decoupled from one another. Furthermore, we also calculated the spin structure factor $S(q_y)$ in Fig.~\ref{Fig5}(d), which displays a mild antiferromagnetic (AFM) coupling along the dimerized chain induced by the spin singlets (again, long-range order is rapidly suppressed). This AFM-AFM coupling is reasonable, considering the facts known about Eq.~(1). {\color{blue}The magnetic coupling in a dimer should be AFM because the large overlap of V-$3d_{\rm xy}$ orbitals establishes AFM coupling in a dimer according to super-exchange ideas. Between neighboring V-V dimers, our DMRG calculations predict a short-range coupling which is also AFM due to the direct V-V magnetic interaction, albeit much weaker.}

\textit{Discussion and perspective.-}
In previous DFT studies~\cite{Tan:prb,Ding:prb}, the FE-I state of VOI$_2$ (without V-V dimerization along the
$b$-axis) was argued to be stable in a long-range FM ordered state with low transition temperature. However, the V-V dimerization was not introduced in their calculations. Our study suggests that the V-V dimerization could suppress the long-range magnetic ordering, as reported in other $S=1/2$ monolayer compounds, such as VSe$_2$ ~\cite{Coelho:jpcc}. It also should be noted that a strong dimerization was also reported experimentally in TaOI$_2$ ($5d^1$)~\cite{Ruck:acc} and MoOCl$_2$ ($4d^2$)~\cite{Wang:prm20}. Hence, our results appear reasonable and in agreement with experiments: the monolayer ground state should be FE-II (with a FE distortion along the $a$-axis,
V-V dimerization along the $b$-axis, and short-range AFM spin order).

In addition, we also calculated the phononic spectrum of the undistorted phase of VOBr$_2$ and VOCl$_2$ (Fig. S15), and the results are similar to those of VOI$_2$. The unstable FE distortion mode was enhanced in VOBr$_2$ and VOCl$_2$, indicating a larger $P$. Furthermore, the FE-II phase is also expected to be the ground state in VOBr$_2$ and VOCl$_2$. Although more work is needed, the physics should be similar to our study in VOI$_2$.

\textit{Conclusion.-}
Here the monolayer compound VOI$_2$ was systematically studied using first-principles DFT and DMRG calculations. A strongly anisotropic metallic band structure was observed in the undistorted phase, suggesting a Peierls instability. In addition, using group symmetry analysis and DFT calculations, we found that the FE-II phase becomes stable at low temperature. This FE-II state can be regarded as arising from the coupling between a ferroelectric distortion along the $a$-axis and V-V dimerization along the $b$-axis. In addition, the ``pseudo Jahn Teller'' effect caused by the coupling between empty V $3d$ and O $2p$ states induces the ferroelectric distortion from the undistorted phase. Furthermore, we also unveiled a robust spin quantum disordered ground state with very short-range antiferromagnetic order, essentially made of spin singlets. Our results successfully produce the expected phase transition induced by the Peierls instability and provide additional insight that should motivate further theoretical and experimental efforts.

\acknowledgments{The work of Y.Z., L.-F.L., A.M. and E.D. is supported by the U.S. Department of Energy (DOE), Office of Science, Basic Energy Sciences (BES), Materials Sciences and Engineering Division. G.A. was partially supported by the scientific Discovery through Advanced Computing (SciDAC) program funded by U.S. DOE, Office of Science, Advanced Scientific Computing Research and BES, Division of Materials Sciences and Engineering. All the calculations were carried out at the Advanced Computing Facility (ACF) of the University of Tennessee Knoxville (UTK).}


\begin{references}
\bibitem{Monceau:ap} P. Monceau, \href{https://doi.org/10.1080/00018732.2012.719674}{Adv. Phys. {\bf 61}, 325 (2012).}
\bibitem{Dagotto:rmp94} E. Dagotto, \href{https://doi.org/10.1103/RevModPhys.66.763}{Rev. Mod. Phys. \textbf{66}, 763 (1994).}
\bibitem{Dagotto:Rmp} E. Dagotto, \href{https://doi.org/10.1103/RevModPhys.85.849}{Rev. Mod. Phys. \textbf{85}, 849 (2013).}
\bibitem{Grioni:JPCM} M. Grioni, S. Pons and E. Frantzeskakis, \href{https://doi.org/10.1088/0953-8984/21/2/023201}{J. Phys.: Condens. Matter {\bf 21}, 023201 (2009).}
\bibitem{Lin:prl} L. F. Lin, Y. Zhang, A. Moreo, E. Dagotto, and S. Dong, \href{https://doi.org/10.1103/PhysRevLett.123.067601}{Phys. Rev. Lett. {\bf 123}, 067601 (2019).}
\bibitem{cu-ladder1} E. Dagotto and T. M. Rice, \href{https://doi.org/10.1126/science.271.5249.618}{Science {\bf 271}, 618 (1996)}.
\bibitem{Zhang:prb19} Y. Zhang, L. F. Lin, A. Moreo, S. Dong, and E. Dagotto \href{https://doi.org/10.1103/PhysRevB.100.184419}{Phys. Rev. B \textbf{100}, 184419 (2019).}
\bibitem{Zhang:prb20-2} Y. Zhang, L. F. Lin, A. Moreo, S. Dong, and E. Dagotto \href{https://journals.aps.org/prb/abstract/10.1103/PhysRevB.101.144417}{Phys. Rev. B \textbf{101}, 144417 (2020).}
\bibitem{gao:prb20} S. Gao, L-F. Lin, A. F. May, B. K. Rai, Q. Zhang, E. Dagotto, A. D. Christianson, and M. B. Stone, \href{https://doi.org/10.1103/PhysRevB.102.220402}{Phys. Rev. B \textbf{102}, 220402(R) (2020).}
\bibitem{cu-ladder2} E. Dagotto, \href{https://doi.org/10.1088/0034-4885/62/11/202}{Rep. Prog. Phys. {\bf 62}, 1525 (1999).}
\bibitem{cu-ladder3} M. Uehara, T. Nagata, J. Akimitsu, H. Takahashi, N. Mori, and K. Kinoshita, \href{https://doi.org/10.1143/JPSJ.65.2764}{J. Phys. Soc. Jpn. {\bf 65}, 2764 (1996).}
\bibitem{Takahashi:Nm} H. Takahashi, A. Sugimoto, Y. Nambu, T. Yamauchi, Y. Hirata, T. Kawakami, M. Avdeev, K. Matsubayashi, F. Du, C. Kawashima, H. Soeda, S. Nakano, Y. Uwatoko, Y. Ueda, T. J. Sato and K. Ohgushi, \href{https://doi.org/10.1038/nmat4351}{Nat. Mater. \textbf{14}, 1008 (2015).}
\bibitem{Zhang:prb17} Y. Zhang, L. F. Lin, J. J. Zhang, E. Dagotto, and S. Dong, \href{https://doi.org/10.1103/PhysRevB.95.115154}{Phys. Rev. B \textbf{95}, 115154 (2017).}
\bibitem{Ying:prb17} J.-J. Ying, H. C. Lei, C. Petrovic, Y.-M. Xiao and V.-V. Struzhkin, \href{https://doi.org/10.1103/PhysRevB.95.241109}{Phys. Rev. B \textbf{95}, 241109(R) (2017).}
\bibitem{Zhang:prb18} Y. Zhang, L. F. Lin, J. J. Zhang, E. Dagotto, and S. Dong, \href{https://doi.org/10.1103/PhysRevB.97.045119}{Phys. Rev. B \textbf{97}, 045119 (2018).}
\bibitem{Gooth:nature} J. Gooth, B. Bradlyn, S. Honnali, C. Schindler, N. Kumar, J. Noky, Y. Qi, C. Shekhar, Y. Sun, Z. Wang, B. A. Bernevig and C. Felser , \href{https://doi.org/10.1038/s41586-019-1630-4}{Nature \textbf{575}, 315 (2019).}
\bibitem{Zhang:prb20-1} Y. Zhang, L. F. Lin, A. Moreo, S. Dong, and E. Dagotto \href{https://doi.org/10.1103/PhysRevB.101.174106}{Phys. Rev. B \textbf{101}, 174106 (2020).}
\bibitem{Lin:prm} L. F. Lin, Y. Zhang, A. Moreo, E. Dagotto, and S. Dong, \href{https://doi.org/10.1103/PhysRevMaterials.3.111401}{Phys. Rev. Mater. \textbf{3}, 111401(R) (2019).}
\bibitem{Choi:prl} Y. J. Choi, H. T. Yi, S. Lee, Q. Huang, V. Kiryukhin, and S.-W. Cheong, \href{https://doi.org/10.1103/PhysRevLett.100.047601}{Phys. Rev. Lett. \textbf{100}, 047601 (2008).}
\bibitem{Tan:prb} H. Tan, M. Li, H. Liu, Z. Liu, Y. Li, and W. Duan, \href{https://doi.org/10.1103/PhysRevB.99.195434}{Phys. Rev. B \textbf{99}, 195434 (2019).}
\bibitem{Ding:prb} N. Ding, J. Chen, S. Dong, and A. Stroppa, \href{https://doi.org/10.1103/PhysRevB.102.165129}{Phys. Rev. B \textbf{102}, 165129 (2020).}
\bibitem{Xu:prl} C. Xu, P. Chen, H. Tan, Y. Yang, H. Xiang, and L. Bellaiche, \href{https://doi.org/10.1103/PhysRevLett.125.037203}{Phys. Rev. Lett. \textbf{125}, 037203 (2020).}
\bibitem{Cohen:nature} R. E. Cohen, \href{https://doi.org/10.1038/358136a0}{Nature \textbf{358}, 136 (1992).}
\bibitem{Cohen:fe} R. E. Cohen and H. Krakauer, \href{https://doi.org/10.1080/00150199208016067}{Ferroelectrics \textbf{136}, 65 (1992).}
\bibitem{Young:prb} D. Hickox-Young, D. Puggioni, and J. M. Rondinelli, \href{https://doi.org/10.1103/PhysRevB.102.014108}{Phys. Rev. B \textbf{102}, 014108 (2020).}
\bibitem{Smaalen:aca} S. van Smaalen, \href{https://doi.org/10.1107/S0108767304025437}{Acta Cryst. A \textbf{61}, 51 (2004).}
\bibitem{Kresse:Prb} G. Kresse and J. Hafner, \href{https://doi.org/10.1103/PhysRevB.47.558}{Phys. Rev. B \textbf{47}, 558 (1993).}
\bibitem{Kresse:Prb96} G.~Kresse and J.~Furthm\"{u}ller, \href{https://doi.org/10.1103/PhysRevB.54.1169}{Phys. Rev. B \textbf{54}, 11169 (1996).}
\bibitem{Blochl:Prb} P. E. Bl\"{o}chl, \href{https://doi.org/10.1103/PhysRevB.50.17953}{Phys. Rev. B \textbf{50}, 17953 (1994).}
\bibitem{Chaput:prb} L. Chaput, A. Togo, I. Tanaka, and G. Hug, \href{https://doi.org/10.1103/PhysRevB.84.094302}{Phys. Rev. B \textbf{84}, 094302 (2011).}
\bibitem{Togo:sm} A. Togo, I. Tanaka, \href{https://doi.org/10.1016/j.scriptamat.2015.07.021}{Scr. Mater. \textbf{108}, 1 (2015).}
\bibitem{Momma:vesta}{\color{blue} K. Momma and F. Izumi, \href{https://doi.org/10.1107/S0021889811038970}{J. Appl. Crystallogr. \textbf{44}, 1272 (2011).}}
\bibitem{Dudarev:prb}{\color{blue}S. L. Dudarev, G. A. Botton, S. Y. Savrasov, C. J. Humphreys, and A. P. Sutton, \href{https://doi.org/10.1103/PhysRevB.57.1505}{Phys. Rev. B \textbf{57}, 1505 (1998).}
\bibitem{Sun:prl}{\color{blue} J. Sun, A. Ruzsinszky, and J. P. Perdew, \href{https://doi.org/10.1103/PhysRevLett.115.036402}{Phys. Rev. Lett. \textbf{115}, 036402 (2015).}}
\bibitem{dmrgcontext}{\color{blue}The Peierls distortion could suppress the FM order since the $90^\circ$ superexchange between neighboring V-I-V bonds would be destroyed. DFT always overestimate ferromagnetic tendencies. They can be a signature of strong quantum fluctuations that may suppress ferromagnetism. Hence, we adopted the advanced many-body DMRG method instead of DFT to discuss the quantum magnetic coupling in this S = 1/2 dimerized chain.}}
\bibitem{Orobengoa:jac} D. Orobengoa, C. Capillas, M. I. Aroyo, and J. M. Perez-Mato, \href{https://doi.org/10.1107/S0021889809028064}{J. Appl. Crystallogr. \textbf{42}, 820 (2009).}
\bibitem{Perez-Mato:aca} J. Perez-Mato, D. Orobengoa, and M. I. Aroyo, \href{https://doi.org/10.1107/S0108767310016247}{Acta Crystallogr. A \textbf{66}, 558 (2010).}
\bibitem{Modecontext} The $\Gamma^{4-}$ mode is a ferroelectric distortion along the $a$-axis, resulting in the FE-I phase, as predicted by previous study ~\cite{Tan:prb}. The $S^{3-}$ and $Y^{1+}$ modes are V-dimerization distortions along the $b$-axis, as shown in Fig.~\ref{Fig2}(d), corresponding to different symmetry breaking patterns. In the $b$-axis, the V-V dimerization would induce a long-short (LS) V-V bonds pattern. In the DIM-I phase, the V-dimerization of each chain remains the same while the V-dimerization is opposite between different VI$_2$ chains in the DIM-II phase.
\bibitem{Supplemental} For more results, see Supplemental Material at \href{http://link.aps.org/supplemental/10.1103/PhysRevB.xx/xxxxxx}{http://link.aps.org/supplemental/10.1103/PhysRevB.xx/xxxxxx.}
\bibitem{Rondinelli:prb} J. M. Rondinelli, A. S. Eidelson, and N. A. Spaldin, \href{https://doi.org/10.1103/PhysRevB.79.205119}{Phys. Rev. B \textbf{79}, 205119 (2009).}
\bibitem{Filippetti:prb} A. Filippetti and N. A. Hill, \href{https://doi.org/10.1103/PhysRevB.65.195120}{Phys. Rev. B \textbf{65}, 195120 (2002).}
\bibitem{pJTcontext}  Furthermore, we also calculated the electronic structures for centrosymmetric and ferroelectric symmetries in the FE-I crystal structure. Similar changes of electronic structures were obtained as shown in Fig.S6.
\bibitem{VCAcontext} Here, the carrier concentrations were considered in the range from $0.25$ holes to $0.25$ electrons per unit cell (undistorted phase) by introducing nearby Ti or Cr on V sites based on the VCA method. Both the in-plane lattice constants and atomic positions were fully relaxed before calculating the frequency of ferroeletric modes under different doping level. To save computing resources, we adopted a $2\times2\times1$ supercell in our calculations for Fig.~\ref{Fig3}(d), since this $2\times2\times1$ superlattice is enough to describe the FE disstortion.
\bibitem{Zhang:prb20} Y. Zhang, L.-F. Lin, W. Hu, A. Moreo, S. Dong, and E. Dagotto, \href{https://doi.org/10.1103/PhysRevB.102.195117}{Phys. Rev. B \textbf{102}, 195117 (2020).}
\bibitem{Bellaiche:Prb} L. Bellaiche and D. Vanderbilt, \href{https://doi.org/10.1103/PhysRevB.61.7877}{Phys. Rev. B \textbf{61}, 7877 (2000).}
\bibitem{Ramer:Prb} N. J. Ramer and A. M. Rappe, \href{https://doi.org/10.1103/PhysRevB.62.R743}{Phys. Rev. B \textbf{62}, R743(R) (2000).}
\bibitem{Zhao:prb18}{\color{blue}H. J. Zhao, A. Filippetti, C. Escorihuela-Sayalero, P. Delugas, E. Canadell, L. Bellaiche, V. Fiorentini, and J. \'I\~niguez, \href{https://doi.org/10.1103/PhysRevB.97.054107}{Phys. Rev. B \textbf{97}, 054107 (2018).}}
\bibitem{FE2context} Although the space group of the FE-II phase is the same as in the FE-I phase, the I atoms occupy two different Wyckoff sites ($2e$ and $2f$) in the FE-I state while the Wyckoff sites of I atoms are in $2e$. Furthermore, the V and O Wyckoff sites are also different (both $2f$ in FE-I phase and $4i$ in FE-II phase), corresponding to different in-plane symmetries.
\bibitem{Savin:Angewandte} A. Savin, O. Jepsen, J. Flad, O.-K. Andersen, H. Preuss, and H. G. von Schnering, \href{https://doi.org/10.1002/anie.199201871}{Angew. Chem. Int. Ed. \textbf{32}, 187 (1992).}
\bibitem{King-Smith:Prb} R. D. King-Smith and D. Vanderbilt, \href{https://doi.org/10.1103/PhysRevB.47.1651}{Phys. Rev. B \textbf{47}, 1651 (1993).}
\bibitem{Resta:Rmp} R. Resta,  \href{https://doi.org/10.1103/RevModPhys.66.899}{Rev. Mod. Phys. \textbf{66}, 899 (1994).}
\bibitem{Mostofi:cpc} A. A. Mostofi, J. R. Yates, Y. S. Lee, I. Souza, D. Vanderbilt, and N. Marzari, \href{https://doi.org/10.1016/j.cpc.2007.11.016}{Phys. Commun. \textbf{178}, 685 (2007).}
\bibitem{white:prl} S. R. White, \href{https://doi.org/10.1103/PhysRevLett.69.2863}{Phys. Rev. Lett. \textbf{69}, 2863 (1992).}
\bibitem{white:prb} S. R. White, \href{https://doi.org/10.1103/PhysRevB.48.10345}{Phys. Rev. B \textbf{48}, 10345  (1993).}
\bibitem{Alvarez:cpc} G. Alvarez, \href{https://doi.org/10.1016/j.cpc.2009.02.016}{Comput. Phys. Commun. \textbf{180}, 1572 (2009).}
\bibitem{Coelho:jpcc} P. M. Coelho, K. N. Cong, M. Bonilla, S. Kolekar, M.-H. Phan, J. Avila, M. C. Asensio, I. I. Oleynik, and M. Batzill, \href{https://doi.org/10.1021/acs.jpcc.9b04281}{J. Phys. Chem. C \textbf{132}, 14089 (2019).}
\bibitem{Ruck:acc} M. Ruck, \href{https://doi.org/10.1107/S0108270195003799}{Acta Cryst. C \textbf{6}, 1960 (1996).}
\bibitem{Wang:prm20} Z. Wang, M. Huang, J. Zhao, C. Chen, H. Huang, X. Wang, P. Liu, J. Wang, J. Xiang, C. Feng, Z. Zhang, X. Cui, Y. Lu, S. A. Yang, and B. Xiang, \href{https://doi.org/10.1103/PhysRevMaterials.4.041001}{Phys. Rev. Mater. \textbf{4},  041001(R)  (2020).}
\end{references}
\end{document}